\documentstyle[twocolumn,psfig]{article}

\makeatletter
\def\sloppy{\tolerance 9999 \hfuzz .5\p@ \vfuzz .5\p@}
\def\section{\@startsection {section}{1}{\z@}{3.5ex plus 1ex minus 
 .2ex}{2.3ex plus .2ex}{\bf}}
\def\fnum@figure{Fig. \thefigure}
\long\def\@makecaption#1#2{
 \footnotesize \vskip 8pt
 \setbox\@tempboxa\hbox{{ #1.} #2}
 \ifdim \wd\@tempboxa >\hsize \unhbox\@tempboxa\par \else \hbox
 to\hsize{\hfil\box\@tempboxa\hfil} 
 \fi}
\makeatother

\begin{document}

 \title{Orbital-ordering-induced anomalous softening of the
 ferromagnetic spin waves in perovskite manganites}
 \author{R. Kajimoto$^{\rm a,*,\dag}$, H. Yoshizawa$^{\rm
 a}$, H. Kawano-Furukawa$^{\rm b,\dag}$, \\
 H. Kuwahara$^{\rm c,\ddag}$, Y. Tomioka$^{\rm c}$, and Y. Tokura$^{\rm c}$ \\
 {\footnotesize\it $^{\rm a}$Neutron Scattering Laboratory, I. S. S. P., 
 Univ. of Tokyo, Tokai, Ibaraki, 319-1106, Japan} \\
 {\footnotesize\it $^{\rm b}$Institute of Physical and Chemical
 Research, Hirosawa 2-1, Wako, Saitama, 351-0198, Japan} \\
 {\footnotesize\it $^{\rm c}$Joint Research Center for Atom Technology, 
 Tsukuba, Ibaraki 305-8562, Japan}}
 \date{\footnotesize (\today)}

\twocolumn[%
\maketitle%
\vspace{-1cm}
\parbox{\textwidth}{%
\vskip 24pt plus 6pt minus 3pt
\hrule height 0.4pt
\vskip 8pt
{\small
\noindent
{\bf Abstract}
\vspace{0.5\baselineskip}

Spin wave excitations were measured in the ferromagnetic phase of
Nd$_{1/2}$Sr$_{1/2}$MnO$_{3}$ by neutron scattering. This compound is
located in proximity to the A-type antiferromagnetic state, and it shows
a clear anisotropy and anomalous softening of the spin wave excitations.
The softening in the ferromagnetic phase is induced by the orbital
ordering.

\vspace{.5\baselineskip}

\vskip 10pt
\hrule height 0.4pt
\vskip 1.5\baselineskip
}}]


\footnotetext[1]{Corresponding author. Fax: +81 3 5978 5325; e-mail:
kaji@phys.ocha.ac.jp}
\footnotetext[2]{Present address: Faculty of Science, Ochanomizu
University, Bunkyo-ku, Tokyo 112-8610, Japan.}
\footnotetext[3]{Present address: Faculty of Science and Technology,
Sophia University, Chiyoda-ku, Tokyo 102-8554, Japan.}


Recent experimental and theoretical studies revealed that the ordering
of the Mn $e_g$ orbitals plays a crucial role to determine physical
properties in the perovskite manganites in addition to the well-known
double exchange interactions. Especially, the strong influence of the
orbital ordering on the magnetism and the transport properties in the
antiferromagnetic (AFM) state are widely recognized. For example, the
metallic A-type AFM state, in which ferromagnetic (FM) planes stack
antiferromagnetically, was suggested to be attributed to the ordering of
$d(x^2-y^2)$ orbitals of Mn ions in the FM planes. The
anisotropic interactions of the $d(x^2-y^2)$ orbitals introduce a
noticeable two dimensional character in the spin
fluctuations\cite{yoshi98}.

On the other hand, the influence of the orbital ordering in the FM or
paramagnetic (PM) state is not clear yet. Accordingly, a neutron
scattering studies on the FM and PM phase of
Pr$_{1/2}$Sr$_{1/2}$MnO$_{3}$ and Nd$_{1/2}$Sr$_{1/2}$MnO$_{3}$, which
are located in proximity to the A-type AFM state, were performed in
order to examine the influence of the orbital ordering on the spin
fluctuations.  The samples are single crystals, and the measurements
were performed on the $(h, l, h)$ scattering plane (in the $Pnma$
setting).  Because the space of this paper is limited, we will only
present the results for Nd$_{1/2}$Sr$_{1/2}$MnO$_{3}$. Similar results
were obtained for Pr$_{1/2}$Sr$_{1/2}$MnO$_{3}$, which will be described
elsewhere.





Nd$_{1/2}$Sr$_{1/2}$MnO$_{3}$ shows the FM transitions at $T_{\rm C} =
250$ K. With decreasing temperature, the magnetic structure is switched
to the CE-type AFM structure at $T_{\rm N}^{\rm CE} = 160$ K as a first
order transition \cite{kawano97}. In addition, we found that the canted
A-type AFM structure appears as a second order transition in the FM
phase below $T_{\rm N}^{\rm A} = 200$ K. Its AFM propagation vector is
(0,1,0) in reciprocal lattice unit. In this paper, we will denote the
direction within or between the FM planes as intraplane or interplane
direction, respectively.

\begin{figure}[htbp]
 \centering
 \psfig{file=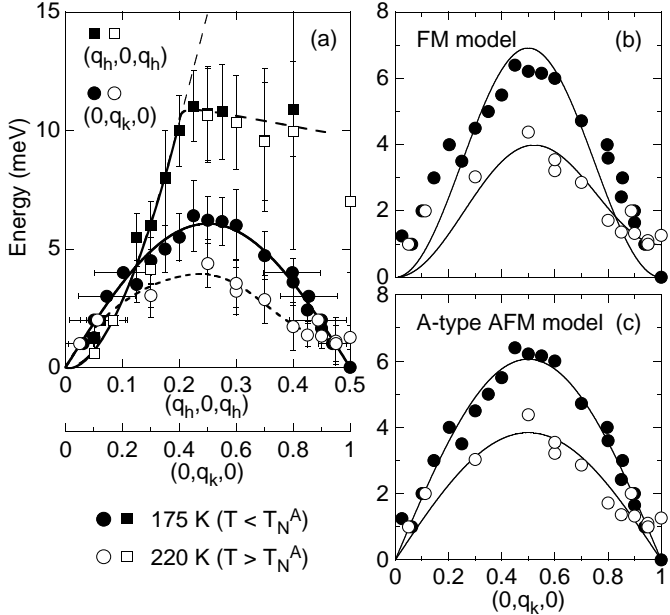,width=\hsize}
 \caption{(a) Spin wave dispersion curves for
 Nd$_{1/2}$Sr$_{1/2}$MnO$_{3}$ in the FM phase. Squares and circles
 indicate data measured along the intraplane direction and those along
 the interplane direction, respectively. Solid symbols are data for 175
 K and open symbols are those for 220 K. (b), (c) Fits to the interplane
 dispersions based on two models (described in the text).}
 \label{spinwave}
\end{figure}

Figure \ref{spinwave} (a) shows the spin wave dispersion curves which
were measured from the zone center to the zone boundary in the FM
phase. Two interesting features were observed. First, the spin wave
excitations show strongly anisotropic behavior, which is similar to the
A-type AFM phase\cite{yoshi98}. Secondly, the dispersion curves exhibit
severe softening near the zone boundary. Especially, the softening of
the dispersion curve along the interplane direction is so strong that
the dispersion resembles that of the A-type AFM state.  We think these
results suggest the existence of the influence of the $d(x^2-y^2)$-type
orbital ordering in the FM phase. In the following, we shall discuss
relations between softening of the interplane spin waves and the orbital
ordering in detail. The discussion on the anisotropy of the spin waves
has been reported elsewhere \cite{kawano_un}.

In order to parametrize the softening phenomenogically, we have fitted
the observed dispersions to the Heisenberg model for ferromagnets,
although the localized spin model is too naive. It is necessary to take
the second nearest neighbor exchanges along the interplane direction
into account to reproduce the softening of the interplane
dispersion. Therefore, we fitted the data by the following equation:
\begin{equation}
 \hbar\omega(q_k)
  = 8J_1S\sin^2\Bigl(\frac{1}{2}\pi q_k\Bigr) + 8J_2S\sin^2(\pi q_k),
  \label{FM_model_eq}
\end{equation}
where $J_1$ and $J_2$ are the exchange integrals for the nearest
neighbors and those for the second nearest neighbors along the
interplane direction. The results are shown in
Fig. \ref{spinwave} (b). Alternatively, we have fitted the data to the
Heisenberg model for antiferromagnet, because the interplane dispersions
resemble the one for the A-type antiferromagnet. Here, we adopted only
the exchanges between the first nearest neighbors. Then, the equation
for the dispersion becomes
\begin{equation}
 \hbar\omega(q_k)
  = -8J_1S\Bigl|\frac{1}{2}\sin(\pi q_k)\Bigr|,
  \label{A-type_model_eq}
\end{equation}
where $J_1<0$. The results of the fitting by Eq. (\ref{A-type_model_eq})
is indicated in Fig. \ref{spinwave} (c). 

Although Eq. (\ref{FM_model_eq}) can reproduce the softening, it fails
to describe the data for the small $q$ region, and $J_1$ becomes very
small compared to $J_2$: $8J_1S=1.0$ meV and $8J_2S=3.5$ meV for 220 K,
while $8J_1S=0$ meV and $8J_2S=6.9$ meV for 175 K. In contrast, the
observed data are rather well described by the AFM dispersion
(Eq. \ref{A-type_model_eq}) except for high-$q$ region at 220 K. In this
case, however, $J_1 < 0$ ($8J_1S=-7.7$ meV for 220 K and $-12$ meV for
175 K) in spite of the fact that the long range FM ordering is formed.

The exchange parameters obtained by these procedures are clearly
anomalous, and it should be attributed to the inadequacy of the
localized Heisenberg spin model. It is clear that theory which takes
into account the relevant microscopic mechanism has to be developed.
We think the fluctuations of the exchange energy induced by the
$d(x^2-y^2)$-type orbital correlation may be responsible for the
softening of the interplane dispersion. This scenario may also explain
the temperature dependence of the interplane dispersion.  As shown in
Fig. \ref{spinwave}, the interplane dispersion has a gap at the zone
boundary above $T_{\rm N}^{\rm A}$, which reflects that the orbital
ordering is not fully stabilized yet. With decreasing temperature, the
orbital ordering becomes stable. As a result, the energy gap decreases
and eventually vanishes in the canted A-type AFM phase, resulting in a
complete A-type AFM dispersion as shown in Fig. \ref{spinwave} (c). The
transition to the A-type AFM states may be regarded as a transition by
the softening of magnons, on the analogy of the structural transition by
the softening of phonons. We would like to note that the similar
softening of the FM spin wave dispersion and its temperature dependence
was reproduced by the recent theoretical calculations which adopted the
orbital correlations \cite{ishihara97,khaliullin00}.


In summary, we found an anomalous softening of the magnon dispersions in
the FM phase of Nd$_{1/2}$Sr$_{1/2}$MnO$_{3}$. The softening is
attributed to the influence of the orbital ordering in the FM phase, and
the FM to A-type AFM transition is induced by the softening of the
magnon.


This work was supported by a Grant-In-Aid for Scientific Research from
the Ministry of Education, Science and Culture, Japan and by the New
Energy and Industrial Technology Development Organization (NEDO) of
Japan.

\end{document}